\documentclass{article}
\usepackage{ijcai22}

\usepackage{times}
\usepackage{soul}
\usepackage{url}
\usepackage[hidelinks]{hyperref}
\usepackage[utf8]{inputenc}
\usepackage[small]{caption}
\usepackage{graphicx}
\usepackage{amsmath}
\usepackage{amsthm}
\usepackage{booktabs}
\usepackage{algorithm}
\usepackage{algorithmic}
\usepackage{multirow}
\title{Deep Reinforcement Learning in Factor Investment }

\author{
Junlin Liu$^1$
\and
Grace Hui Yang$^2$
\affiliations
$^{1,2}$Georgetown University\\
\emails
jl2748@georgetown.edu,
Grace.Yang@georgetown.edu,
}

\begin{document}

\maketitle

\begin{abstract}
Deep reinforcement learning (DRL) has shown promise in trade execution, yet its use in low-frequency factor portfolio construction remains under-explored. A key obstacle is the high-dimensional, unbalanced state space created by stocks that enter and exit the investable universe. We introduce Conditional Auto-encoded Factor-based Portfolio Optimisation (CAFPO), which compresses stock-level returns into a small set of latent factors conditioned on 94 firm-specific characteristics. The factors feed a DRL agent—implemented with both PPO and DDPG—to generate continuous long–short weights. On 20 years of U.S. equity data (2000–2020), CAFPO outperforms equal-weight, value-weight, Markowitz (historical \& factor), vanilla DRL, and Fama–French-driven DRL, delivering a 24.6\% compound return and a Sharpe ratio of 0.94 out of sample. SHAP analysis further reveals economically intuitive factor attributions. Our results demonstrate that factor-aware representation learning can make DRL practical for institutional, low-turnover portfolio management.
\end{abstract}

\section{Introduction}

DRL is found an outstanding tool for trading from an array of experiments\cite{briola_deep_2021,liu_adaptive_2020,si_multi-objective_2017,yu_model-based_2019,liang_adversarial_2018}. However, their application to portfolio optimization remains unclear and have challenges.First, a portfolio tends to have a long holding horizon, ranging from months to years. During the holding period, stocks in the portfolio are usually not fixed. Undergoing corporate actions like mergers, acquisitions, and bankruptcies delist securities from stock exchanges, stocks can simply be unavailable. Portfolio manager’s stock selection criteria can also remove or add stocks to the portfolio based on their analysis from time to time. If stock prices or returns were set as state space, like existing studies do, there would be many missing values at a given period. Secondly, institutional portfolio usually has lower rebalance frequency that can greatly reduce available time series observations from thousands of trade events per day to around a hundred observations per year. While a portfolio manager can easily hold a portfolio with more than 100 stocks, assigning weights to them given such a small training sample is an intractable problem. 

To overcome the challenges, we adopt an increasingly popular multifactor framework. The idea is to reduce state space dimension and use DRL to learn the risk-payout relationship between factors and stocks. Factors is popular horse power in dimension reduction. In a variance-covariance matrix estimation for N companies, for example, to directly compute the matrix, we will need to estimate $N \times (N+1)/2$ parameters, which becomes increasing difficult when N grows large. With multifactor model, we will only need to estimate $N \times K$(stocks to factor exposre matrix) plus $K \times K$(factor return covaraince matrix).

The idea of factor-based state space emerges from financial economics that portfolio equity returns can be explained by a set of financial and economic variables. They are persistent and condensed representations of a portfolio, thus providing a desirable state space. One of the most notable factor models is from \cite{fama_common_1993} showing that the cross-section of average returns on U.S. common stocks can be described by Market, Size, and Value factors, which together explained 89-96\% of a portfolio’s return \cite{bhatnagar_capital_2012}. They further extend the factor model to Fama-French Five-Factor (FF5) by adding Profitability and Investment factors \cite{fama_dissecting_2007,lewellen_cross_2014}. This is usually represented as a regression:
\begin{equation}
    \begin{split}
        r^{e}_{i,t} &= a_i + \beta_{M,i}Market_t + \beta_{S,i}Size_t + \beta_{V,i}Value_t \\
        &+  \beta_{P,i}Profitability_t + \beta_{I,i}Investment_t + e_{it}
    \end{split}
\end{equation}%

In the empirical analysis, we use FF5 factor-based model as a comparison method. Factor models also create a new investment approach that involves targeting specific drivers of returns across assets \cite{noauthor_factors_nodate}. Timely exposure to factors leads to strong portfolio returns. Hence, if a DRL agent learns how to allocate factor exposures for future rewards and learns the relationship between factors and stocks, they will know how to better allocate stock weights.

In light of this idea, we let DRL agents learn from the factors. Up to the point of FF5, factor models’ goal is to explain equity returns using a linear combination of a selected number of factors. For this model to be incorporated into DRL agents, it has two shortcomings. While DRL is able to learn complex relationships between features and rewards, the linearity assumption imposed on factor models hinders neural networks. Secondly, if we allow the collection of factors grows large into high-dimensional, learning can become intractable. To overcome the two challenges, we employ a dimension-reduction model in deep learning, known as autoencoder, to compress high-dimensional stock returns to latent factors while allowing nonlinear and interactive relationships. Furthermore, in order to improve autoencoder’s estimates and provide direct economic interpretations like other factor models, we condition autoencoded latent factors on firm-specific covariates \cite{gu_autoencoder_nodate}. Combining their advantages, we propose a Conditional-Autoencoded Factor-based Portfolio Optimization method (CAFPO). We conduct experiments over 20 years of historical monthly stock returns from US-based exchanges, including NYSE, NASDAQ, and AMEX. CAFPO significantly improves portfolio performance upon Markowitz methods and other learning-based methods.

In finance, the ability to explain a model is extremely important. Our proposed factor-based DRL methods provide a factor investing interpretation to users, transparentizing the optimization process and engendering appropriate trust. We use SHAP value to evaluate factor contributions to provide stock and portfolio-level insights. This paper makes several important contributions:
\begin{itemize}
    \item We propose a practical framework that enables the DRL method to learn from factors and perform low-frequency equity portfolio optimization tasks. This framework extends to CAFPO which significantly outperforms traditional optimization and other learning-based methods.
    \item Our work provides a direct connection between DRL-based trading and factor models in financial economics, providing a new possibility for future research in AI Finance.
    \item We propose to use SHAP value to attribute portfolio decisions to factors, supporting understanding of the optimization process.
\end{itemize}

\section{Related Works}

\paragraph{DRL-based Trading.} Quantitative finance field has adopted DRL in algorithmic trading. Some works design agents to output discrete actions: hold, selling and buying \cite{li_empirical_2019,liu_practical_2022,yang_deep_2020}. This approach focuses on market timing to learn when is the best time to buy low and sell high, but leaves out combining properties of individual assets. Portfolio optimization, on the other hand, is normally seen as a continuous action space problem when capital is significantly larger than the stock price. Researchers have introduced a DRL framework that learns from price data and outputs a continuous portfolio vector \cite{jiang_deep_2017}. This approach achieves outstanding results on high and medium-frequency trading in cryptocurrency \cite{jiang_cryptocurrency_2017}, and Chinese markets \cite{liang_adversarial_2018}. The latest work has shown improved performance after adding technical and volume features for better state description \cite{ren_use_2021}. As mentioned in the last section, there are other works that utilize DRL on portfolio optimization, but they rely on a high and medium-frequency setting with a fixed universe and/or abundant high-frequency data \cite{wang_deeptrader_2021,soleymani_financial_2020,xucheng_portfolio_2019}. Despite the difference in portfolio setting, the previous works establish that DRL-based methods can achieve outstanding trading results from learning a continuous portfolio vector and that adding informative features can enhance portfolio performance. 

\paragraph{Factor Models.} 
Factor models are crystalized stock information that researchers have identified and processed. They were first proposed by \cite{sharpe_capital_1964,lintner_valuation_1965} that the cross-section of average returns on U.S. common stocks can be explained by a single market factor. The market factor is the excess return of the overall US market. From single-factor to three-factor \cite{fama_common_1993}, and to five-factor, factor models keep evolving as researchers gain more understanding about the market. Institutional investors like BlackRock and Invesco rely on factor models to make investment decisions \cite{noauthor_factors_nodate,noauthor_invesco_nodate}. The latest works incorporate machine learning and deep learning into factor models to predict stock returns and explain market dynamics \cite{gu_autoencoder_nodate,gu_empirical_2019}. Our work aims to demonstrate that DRL, as a sub-field of machine learning, can be combined with factor models to perform low-frequency equity management.

\begin{figure*}
  \includegraphics[width=\textwidth,height=7.5cm]{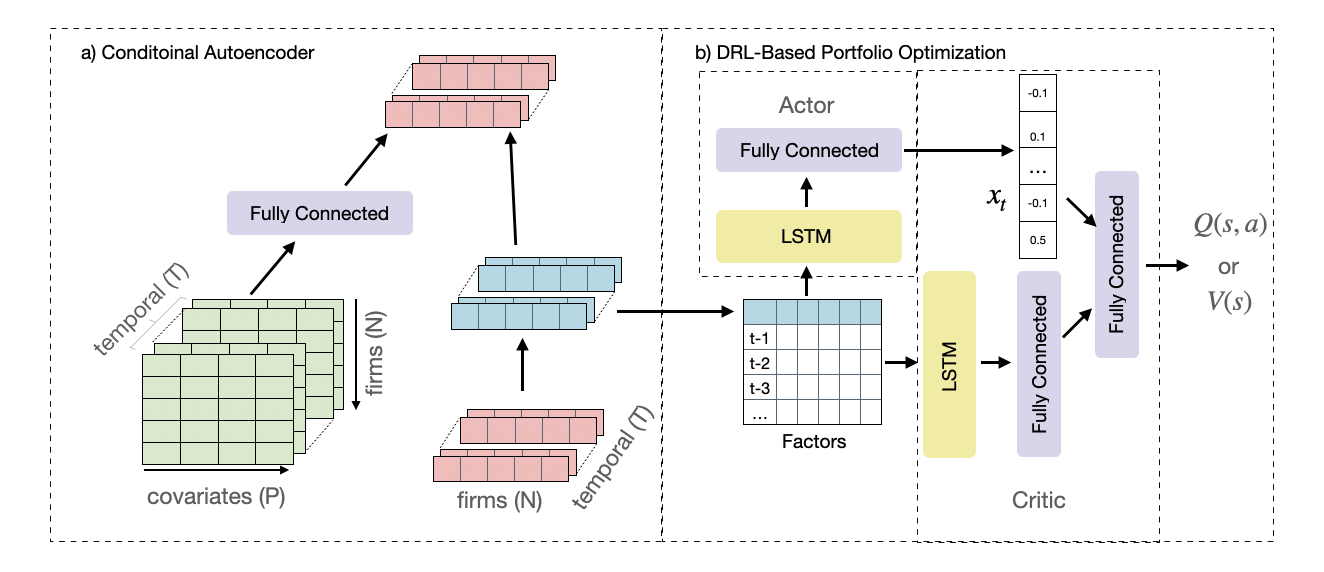}
  \caption{Overarching process.}
\end{figure*}

\section{Problem Setting}

A portfolio is a collection of multiple financial assets. In this paper it represents a collection of stocks. Consider N stocks in a portfolio. Let $v_t = [v_{1,t} ... v_{n,t}]^T \in R^N$ denotes the asset price vector, and $h_t = [h_{1,t} ... h_{n,t}]^T \in R^N$ denotes the number of holding shares. Hence, we have portfolio value $W_t = v_t^Th_t \in R$, single stock returns $r_{i,t} = \frac{v_{i,t} - v_{i,t-1}}{v_{i,t-1}}$ of asset i from time $t-1$ to $t$. Lastly, the portfolio return is defined as $r_{p,t} \in R$ from time t-1 to t equals $\frac{W_t - W_{t-1}}{W_{t-1}} = \sum_{i=1}^Nr_{i,t}x_{i,t} = r_t^Tx_t$.

Portfolio optimization determines the optimal percentage of holdings $x_{i,t} = \frac{h_{i,t}v_{i,t}}{W_t}$ so that the portfolio performance is maximized. This paper adopts three commonly used measures to assess portfolio performance: compound return, variance-adjusted return (Sharpe Ratio), and drawdown-adjusted return (Sterling Ratio). They are defined as:
\begin{equation}
        \text{Compound Return} = \prod_{t=1}^T (1 + r_{p,t}) - 1
\end{equation}

\begin{equation}
        \text{Sharpe Ratio} = \frac{\text{Average}(r_{p,t})}{\text{Standard Deviation}(r_{p,t})}
\end{equation}

\begin{equation}
        \text{Sterling Ratio} = \frac{\text{Average}(r_{p,t})}{\text{Maximum Draw-Down}}
\end{equation}

\section{Methodology}

This section introduces the implementation details of CAFPO. We will describe our design of state space, action space, reward functions, and agent algorithms.

\subsection{State Space}

To overcome the challenges of low-frequency market data, we form a consistent and low-dimensional state space by applying a dimension-reduction model in deep learning, known as autoencoder. An autoencoder has two main parts: an encoder that maps original data to dense representation, and a decoder that reconstructs the original data \cite{bank_autoencoders_2021}. We aim to reconstruct equity returns $r_t$. To improve estimates of the latent factors, we allow the autoencoder to condition on firm-specific covariates that have return predictability to guide the factor formation. Conditional autoencoder is described in Fig 1 a). At the highest level, the unsupervised model has the form \cite{gu_autoencoder_nodate}:
\begin{align}
    r_{i,t} = \beta'_{i,t-1}f^{CA}_t + u_{i,t}
\end{align}
where $\beta'_{i,t-1} \in R^{N \times K}$ is the output of covariates network, the factor loading matrix, $f^{CA}_t \in R^{K}$ is the factors we aim to estimate, and $u_{i,t} \in R^{N}$ is an idiosyncratic error. K is a hyperparameter that stands for the number of factors. The reconstructed returns equal to cross-product of $\beta'_{i,t-1}$ and $f^{CA}_t$.

\subsubsection{Covariates Network (Left-side of Figure 1a.)}

The neural network formulation for the covariates is:
\begin{align}
    z^{L_\beta}_{i,t-1} = LeakyReLU(b + Wz_{i,t-1}) \\
    \beta_{i,t-1} = b^{L_\beta} + W^{L_\beta}z^{L_\beta}_{i,t-1}
\end{align}
where $W \in R^{H \times P}, b \in R^H$  are weight parameter matrix and bias parameter vector, $W^{L_\beta} \in R^{K \times H}, b^{L_\beta} \in R^K$ are factor exposures layer weight parameter matrix and bias parameter vector, and $z_{i,t-1} \in R^{N \times P}$ is the lagged firm characteristics data. P is the number of covariates. H is a hyperparameter for hidden layer dimension. 

\subsubsection{Factor Network (Right-side of Figure 1a.)}

The neural network formulation for factors is:
\begin{align}
    f^{CA}_{t} = \Tilde{b} + \Tilde{W}r_{t}
\end{align}
where $\Tilde{W}$ and $\Tilde{b}$ are weight parameters matrix and bias parameter for equity returns.

To optimize the autoencoder neural network in Figure 1a., we choose Adam to minimize the mean squared error difference between actual returns $r_{i,t}$ and modeled returns $r'_{i,t}$.

The overall CAFPO state space $F^{CA} = (f^{CA}_t)^T_{t=0} = (f^{CA}_1, f^{CA}_2, ..., f^{CA}_T)$ is a time series sequence of factors. At the broadest level, factors are any low-dimensional temporal representations of equity returns. They can be latent factors generated from modeling like conditional autoencoders mentioned above or observable factors like Size and Value found by \cite{fama_common_1993}. Latent factors are more flexible to model, whereas observable factors provide economic interpretation to the portfolio decisions. Note that equation (8) is simply a linear layer from equity returns to factors. If P = K, $f^{CA}_{t}$ inherits the economic interpretation of the covariates. For example, a market capitalization variable would directly transform into a market capitalization factor, which is also known as the Size factor. We will later discuss the usefulness of an interpretable factor-based state space in section 6.

Our proposed factor-based state space accomplishes three things at once. First, it collapses N-dimensional equity returns to K factors. Secondly, it sidesteps the issues of missing returns since factors at each point in time are guaranteed non-missing. Thirdly, it provides more predictive features on future portfolio performance.

\subsection{Action Space}

Our action represents a portfolio weight vector at time t \cite{jiang_deep_2017}, where i represents the ith stock in the universe. Consider a long-short portfolio, the action is defined as
\begin{equation}
    \begin{split}
        a_t &= x_t = [x_{1,t}, x_{2,t},...,x_{n,t}] \in R^N \\ 
        \sum_ix_i^+ &= 1 \quad \sum_ix_i^-1 = -1, \quad -1 \leq x_{i} \leq 1
    \end{split}
\end{equation}

\subsection{Reward Functions}

Reward $R_t$ is a scalar value, which indicates how well the portfolio is doing at time t. Note that the value of Reward $R_t$ is different from stock returns $r_{i,t}$ and portfolio return $r_{p,t}$. From the Problem Setting section, we use compound return, Sharpe Ratio, and Sterling Ratio as performance measures. However, none of the performance measures is additive. For proper on-line learning, we instead compute log returns, differential Sharpe ratio, and differential Downside Deviation ratio as rewards.

\subsubsection{Log Returns}

Taking natural log of the first term of Equation 2 would give us
\begin{equation}
    \begin{split}
        ln(\prod_{t=1}^T (1 + r_{p,t})) &= \sum_{t=1}^T ln(1 + r_{p,t})\\
        &= ln(\frac{W_2}{W_1}) + ln(\frac{W_3}{W_2}) + ... + ln(\frac{W_T}{W_{T-1}}) \\
        &= ln(W_T) - ln(W_1)
    \end{split}
\end{equation}
As we can see, the difference in log returns between initial and final periods is merely the sum of logarithmic returns for each period. 

\subsubsection{Differential Sharpe Ratio}

We expanded Equation 3 to first order in the adaption rate $\eta$
\begin{equation}
    Sharpe_t \approx Sharpe_{t-1} + \eta \frac{dSharpe_t}{d\eta}|_{\eta=0} + O(\eta^2)
\end{equation}

Since only the first order term in Equation (11) depends on the portfolio return $r_{p,t}$, we define the differential Sharpe ratio as
\begin{align}
\frac{B_{t-1}\Delta A_t-A_{t-1}\Delta B_t/2}{(B_{t-1}-A_{t-1}^2)^{3/2}}
\end{align}

where $A_t$ and $B_t$ are exponential moving estimates of the first and second moments of $r_{p,t}$. They have the form
\begin{equation}
    \begin{split}
        A_t=\eta R_t+(1-\eta)A_{t-1}=A_{t-1}+\eta\Delta A_t \\
        B_t=\eta R_t^2 +(1-\eta)B_{t-1}=B_{t-1}+\eta\Delta B_t\\
    \end{split}
\end{equation}

\subsubsection{Differential Downside Deviation Ratio}

Minimizing maximum-drawdown for Sterling ratio is cumbersome, so we instead optimize Downside Deviation Ratio (DDR)
\begin{equation}
        \text{DDR} = \frac{\text{Average}(r_{p,t})}{(\frac{1}{T}\sum_{t=1}^Tmin(r_{p,t}, 0)^2)^{1/2}}
\end{equation}

In a similar manner to the development of Differential Sharpe Ratio, we expand it to first order in the adaptation rate $\eta$
\begin{equation}
    DDR_t \approx DDR_{t-1} + \eta \frac{dDDR_t}{d\eta}|_{\eta=0} + O(\eta^2)
\end{equation}

Since only the first order term depends on $r_{p,t}$. We define the differential downside deviation ratio as
\begin{equation}
    \begin{split}
        &\frac{r_{p,t} - \frac{1}{2}A_{t-1}}{DD_{t-1}}, \quad r_{p,t} > 0 \\
        &\frac{DD_{t-1}^2 * (r_{p,t} - \frac{1}{2}A_{t-1}) - \frac{1}{2}A_{t-1}r_{p,t}^2}{DD_{t-1}^3}, \quad r_{p,t} \leq 0 \\
    \end{split}
\end{equation}

where $A_t$ and $DD^2_t$ are exponential moving averages of returns and of the squared drawdown,
\begin{equation}
    \begin{split}
        A_t&=A_{t-1}+\eta(r_{p,t} - A_{t-1}) \\
        DD_t^2&=DD_{t-1}^2 + \eta(min\{r_{p,t}, 0\}^2 - DD_{t-1}^2)\\
    \end{split}
\end{equation}

$\eta$ is a hyperparameter, which controls the magnitude of the influence of $r_{p,t}$ on overall portfolio performance. To sum up, we use Log Return, Differential Sharpe Ratio, and Differential DDR as reward functions to optimize portfolio measures in Equations (2), (3), and (4). They provide better interpretability and represent the influence of the portfolio return $r_{p,t}$ for each portfolio measure realized a time step t, thus enabling efficient on-line optimization \cite{moody_learning_2001,liu_adaptive_2020}. In the empirical study, we examine the effects of the three reward functions.

\begin{table*}[t]
\centering
\begin{tabular}{cccc}
\hline
Method & Compound Returns & Sharpe Ratio & Sterling Ratio \\
\hline
Equal-Weight & -42.62\% & -1.16 & -0.04 \\
Value-Weight & -31.36\% & -0.57 & -0.03 \\
Markowitz with   Historical   Estimate & -0.72\% & -0.92 & -0.02 \\
Markowitz with Factor Estimate & -0.31\% & -1.51 & -0.04 \\
Vanilla DRL & 1.45\% & 0.16 & 0.02 \\
FFPO & 2.87\% & 0.24 & 0.01 \\
\textbf{CAFPO} & \textbf{24.58\%} & \textbf{0.94} & \textbf{0.07} \\
\hline
\end{tabular}
\caption{Out-of-sample performance, from January 2000 to December 2020. Learning-based results reported above uses PPO as base learner and log return as reward functions.}
\end{table*}

\subsection{Agent Algorithm}

Several continuous control methods have been proposed, such as policy gradient, dual DQN, Deep Deterministic Policy Gradient (DDPG), and Proximal Policy Optimization (PPO). We use the latter two actor-critic based algorithms to implement our trading agent, representing deterministic and stochastic policy methods. They are efficient for handling high-dimensional and continuous action space \cite{schulman_proximal_2017,lillicrap_continuous_2019}. As shown in Figure 1b., the DRL process maintains an actor-network and critic-network. The actor-network maps states to actions where states are factors and actions are portfolio vectors, and the critic-network outputs the value of state $V(s)$ for PPO and outputs the value of action under that state $Q(s,a)$ for DDPG. 

To have an apple-to-apple comparison between the two DRL methods, we give DDPG and PPO the same network architectures, using LSTM to capture the temporal information in latent factors \cite{wei_capturing_2022} followed by feed-forward layers.

\section{Empirical Study of US Equities}

\subsection{Data}

This paper obtains monthly total individual equity returns from CRSP for all stocks listed in the NYSE, AMEX, and NASDAQ. Our sample begins in December 1979 and ends in December 2020, totaling 40 years. US-based common stocks are identified as the subset of these securities that have a share code value of either 10 or 11. To adjust returns, this paper follows the standard process described in \cite{alma991002963379705251}: 1) Calibrate delisting stock returns and 2) Subtract monthly risk-free rate from adjusted returns to get excess returns. 

For covariates in the conditional autoencoder, we build 94 stock-level variables including financial ratios and fundamental data. The detailed list of variables can be found in \cite{gu_empirical_2019}. To avoid forward-looking bias, we lag firm variables by the schedule (monthly variables lag 1-month, quarterly variables lag 4-month, annual variables lag 6-month), to ensure the latest data were published at least for a month. For missing attributes, we replace them with the cross-sectional median of that attribute during that month. Lastly, we rank-normalize the 94 characteristics into the interval (-1,1) for each month \cite{gu_empirical_2019}.

\subsection{Baseline Methods}

Our considered baseline methods include four categories: 1.Equal-weighted and value-weighted portfolios as the weakest baseline without any optimization 2.Portfolio with Markowitz optimization approach 3.DRL portfolio without factors 4. DRL portfolio with FF5 factors.

Equal-weighted portfolio simply assigns the same weight to all the stocks on the same long/short side. Value-weighted portfolio assigns weight proportional to equity's market capitalization. They give a baseline when no optimization is applied. Markowitz optimization models derive the optimal portfolio by maximizing Sharpe Ratio with estimated returns and covariance matrix. We consider two approaches to estimate expected returns $\mu$ and covariance matrix $V$. A naïve approach is to use the historical sample mean and sample covariance as estimates(historical):
\begin{equation}
    \hat{\mu} := \frac{1}{T}\sum_{t=1}^Tr_t, \quad \hat{V} := \frac{1}{T-1}\sum_{t=1}^T(r_t - \hat{\mu})(r_t - \hat{\mu})^T
\end{equation}

Another approach is to use lagged FF5 as estimates (factor) \cite{ross_arbitrage_1976,fan_high_2008}:
\begin{equation}
    \begin{split}
    r_{t} = \alpha &+ Df^{FF5}_{t-1} + \epsilon \\
    \hat{\mu}_f :=  \alpha + DE[f^{FF5}_{t-1}]&, \quad \hat{V}_f := DV_f^{FF5}D^T + V_{\epsilon}
    \end{split}
\end{equation}

where, $\alpha$ is the intercept, $\epsilon$ is the error term, D is the loading matrix, $f^{FF}_{t-1}$ is the lagged Fama-French five factors described in Equation (1), and $V_f^{FF5}$ is the covariance matrix of returns of the risk factors. 

We also compare our proposed method with the vanilla DRL method, which uses stock returns of the latest twelve months as state space. For stocks with missing returns, we fill them with zeros.

Lastly, to demonstrate the efficiency of the conditional autoencoded factors, we build another factor-based method as a baseline by replacing $f^{CA}_t$ with the Fama-French five factors $f^{FF}_t$ (FFPO). We obtain the Fama-French five factors from Kenneth R. French - Data Library Website. The conditional autoencoder process in Figure 1a. is replaced with the downloaded factors.

\begin{table*}[t]
\centering
\begin{tabular}{c|ccc|ccc|ccc|}
\cline{2-10}
\multirow{2}{*}{Algorithm} & \multicolumn{3}{c|}{Log Returns} & \multicolumn{3}{c|}{Differential Sharpe Ratio} & \multicolumn{3}{c|}{Differential DDR} \\  \cline{2-10}
 & Returns & Sharpe & Sterling & Returns & Sharpe & Sterling & Returns & Sharpe & Sterling \\ \hline
PPO & 24.58\% & 0.94 & 0.07 & 6.29\% & 0.41 & 0.02 & 22.62\% & 0.85 & 0.05 \\
DDPG & 6.25\% & 0.42 & 0.02 & -0.89\% & 0.03 & 0.00 & 8.23\% & 0.62 & 0.04 \\ \hline
\end{tabular}
\caption{Effects of Algorithms and Reward Functions with CAFPO}
\end{table*}

\begin{figure}
  \includegraphics[width=\linewidth,height=5cm]{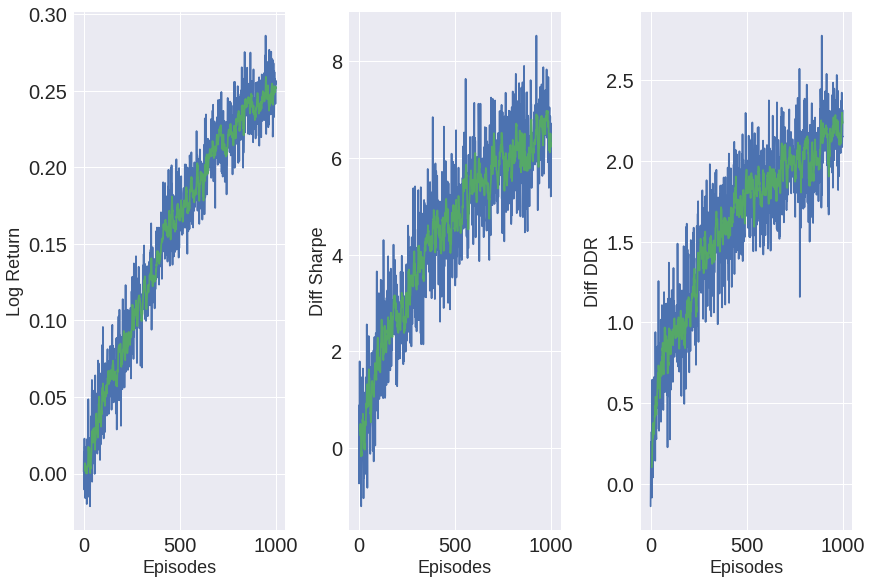}
  \caption{Average total reward (Blue) vs episodes, and moving average (Green)}
\end{figure}

\subsection{Training}

To isolate stock selection effect, we experiment with a simple strategy to hold the 200 stocks with the largest market capitalization of that year. The goal is to assign optimal weights for the 200 stocks in the portfolio.

This paper applies cross-validation in a rolling fashion. Because neural network models are computationally expensive, I avoid recursively retraining the models each month. Instead, I retrain the autoencoder and the agent once every year and use the first 10 years (1989 - 1999) as the validation sample to choose hyperparameters. Each time I retrain, the testing sample rolls forward by 1 year. The testing sample is always 1 year after the training sample. After testing, the training sample rolls forward by 1 year. The training sample is always 10 years before the testing sample. Thus, both training and testing samples are always kept a fixed size of 10 years and 1 year. Our testing sample starts in January 2000.

With neural networks' enhanced flexibility, they are prone to overfitting. In our empirical study, we use five random seeds to initialize DRL's neural networks and construct portfolio weights by averaging the portfolio weights from all networks.

\subsection{Main Results}

Table 1 shows the Out-of-sample backtest results from January 2000 to December 2020. There are several observations: 1) Learning-based methods can indeed outperform traditional optimization methods and naive baselines, generating positive returns for a long-short portfolio. 2) Factor-based methods FFPO and CAFPO improve upon naive DRL-based method, showing the efficiency of factor-based models. 3) CAFPO performs significantly better than FFPO, achieving 24.58\% return. We attribute the performance gains to two reasons. First, while Fama-French factors are built for the entire US equity market, conditional autoencoder factors are built solely from stocks in the portfolio, better capturing the portfolio's return profile. Secondly, conditional autoencoded factors are generated with more information by including significantly more firm-specific characteristics (94 vs 5), and allowing nonlinear and interactive effects.

\subsection{Effects of Algorithms and Rewards}

We further experiment DRL methods with different algorithms and reward functions, as presented in Table 2. We find that while DDPG also generates better performance than Markowitz methods and vanilla DRL method, PPO still yields significantly better results. This is in agreement with the previous work from \cite{yang_deep_2020}. For reward functions, we see that log-return as the reward function constantly forms the best agent regardless of which evaluation metrics are used. This result is in line with our expectations. Neural network applies non-linearities to weighted sums of input features, so they work best when input features are directly linked with rewards.  $f^{CA}_t$ in the state space can be interpreted as returns of non-linearly weighted factor portfolios. With this state space, DRL methods would generate the best result when the reward function is also return. Moreover, as shown in Figure 2., the agent with log-return as the reward function has the sharpest learning curve, while the other two learning curves slow down more quickly and are more volatile. It corroborates our findings that log-return is a more effective reward function than differential Sharpe ratio and differential DDR.

\begin{figure}
  \includegraphics[width=\linewidth,height=5cm]{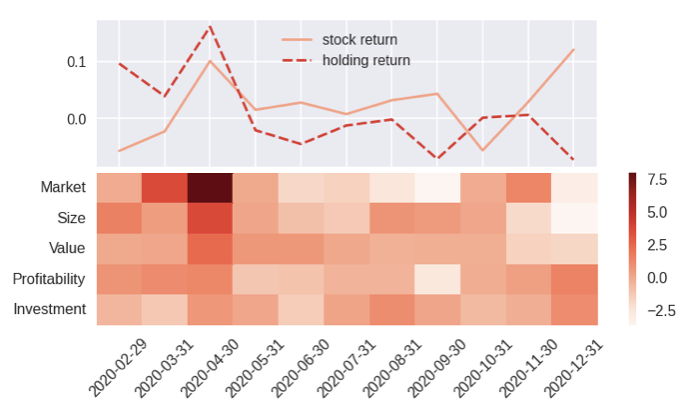}
  \caption{(top) Oracle Corp's stock returns and returns generated from the portfolio by trading this stock. (bottom) factor contribution at each period estimated by SHAP values generated from the actor neural network.}
\end{figure}

\section{Interpretable Investment Decision}

Investments are expensive decisions, especially for financial institutions with millions of dollars under management. Lack of interpretability often arises in this field, preventing advanced but complex models from being used, as explanations are usually mandated. Our proposed factor-based methods have great explanatory strength. As large institutions have adopted factor-based portfolio analysis for years \cite{noauthor_factors_nodate,noauthor_invesco_nodate}, they understand and are comfortable with factor-based decisions. For example, a portfolio manager may instruct to allocate the Size factor a large weight and the Value factor a small weight for the next month. Their employees will then optimize stock weights that meet the factor criteria. Our proposed method fits into the same paradigm that generates stock weights based solely on factors, building factor-driven portfolios. 

We use DeepSHAP to measure each factor’s contribution to the final portfolio. DeepSHAP estimates SHAP value for neural network models to explain the contribution of each input value towards the prediction target, which is optimized stock weights in our case. At any given time, DeepSHAP generates a (look back window M) $\times$ (number of factors K) SHAP matrix for each stock. This gives us the flexibility to calculate factor effects on stock and portfolio-level. 

Stock-level contribution is defined as:

\begin{equation}
	SHAP_{i,t} = \frac{1}{M}\sum_j^MSHAP_{i,j,t}
\end{equation}

Figure 3. plots the stock return, portfolio return from holding this stock, and factor contribution of Oracle Corp during 2020. In this chart, factor contributions are extracted from FFPO with PPO as the base learner. In April 2020, our portfolio put much positive weight on factors, and this generate great portfolio return when Oracle’s return was positive that month. We can see that the Market factor contributes the most during that month. In fact, April 2020 marked the top three best monthly performances for S\&P500 since at least 1950, as governments responded to the COVID-19 pandemic with massive and unprecedented fiscal and monetary stimulus \cite{marketinsite_april_nodate}. It means the DRL method acts perfectly by putting weight on the Market factor. In December 2020, our portfolio puts much weight shorting the factors which hurt our performance when Oracle’s price went up. We can see that the Size and Market factor contribute the most to our short positions. In December 2020, the stock market jumped to record highs. Only the small caps of the Russell 2000 failed to post a gain \cite{management_markets_2021}. Therefore, the DRL method is only correct at shorting the Size factor since small-cap companies went down. However, shorting the Market factor drags the portfolio to a negative return that month.

Portfolio-level contribution is defined as:

\begin{equation}
	SHAP^p_{t} = \frac{1}{N}\frac{1}{M}\sum_i^N\sum_j^M|SHAP_{i,j,t}|
\end{equation}

Here we use absolute value since positive and negative values have the same effect on a long-short portfolio. We then normalize $SHAP^p_{t}$ to get a relative contribution. Figure 4. is an example comparing factors’ contribution and their actual performance during 2020. If a factor has high performance at a time, then the agent should assign a high weight to that factor to maximize return. Likewise, if a factor has low performance, then the agent should avoid assigning any weight. Factor contribution can be calculated at the beginning of a month. It allows a portfolio manager to know how the DRL agent will allocate factor weights for the next period. 

Stock and portfolio-level analytics provide important transparency into how DRL-based methods optimize portfolio, engendering user trust and providing additional insights into how a model may be improved.

\begin{figure}
  \includegraphics[width=\linewidth,height=4cm]{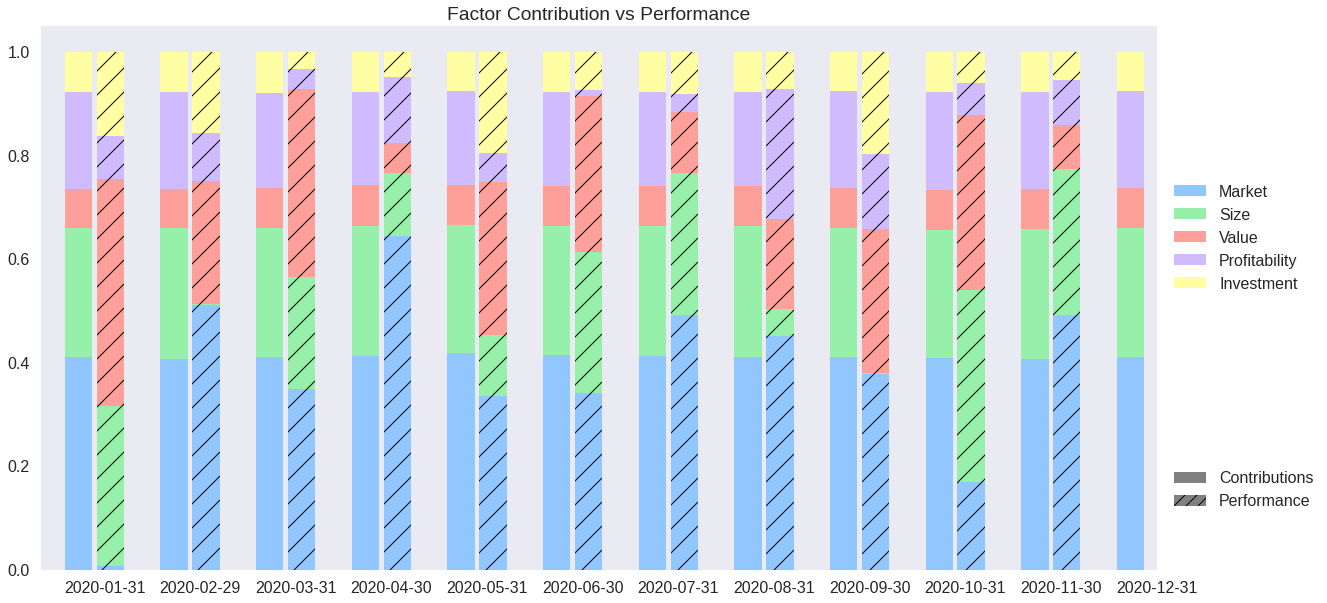}
  \caption{Portfolio-level factor performance vs factor contribution at different periods, estimated by SHAP values generated from the actor neural network.}
\end{figure}

\section{Conclusion}

This paper focuses on low-frequency equity portfolio optimization. We propose a factor-based DRL solution that leverages conditional autoencoder to provide a more meaningful state space. The proposed CAFPO method achieves outstanding results. In the asset management industry, where explanations are mandated, factor-based models represent a promising direction that allows well-known finance concepts to explain complex and advanced trading agents. 

\bibliographystyle{named}
\bibliography{ijcai22}

\end{document}